\documentclass{pasj00}

\begin{document}
\SetRunningHead{Sawada-Satoh et al.}{Internal Motion of 6.7-GHz Methanol Maser in S269}
\Received{2013/01/09}
\Accepted{2013/03/25}

\title{Internal Motion of 6.7-GHz Methanol Masers 
in H\emissiontype{II} Region S269}



%
 \author{%
   Satoko \textsc{Sawada-Satoh}\altaffilmark{1} 
   Kenta \textsc{Fujisawa}\altaffilmark{2}
   Koichiro \textsc{Sugiyama}\altaffilmark{2} 
   Kiyoaki \textsc{Wajima}\altaffilmark{2,3}
   Mareki \textsc{Honma}\altaffilmark{4,5}}
 \altaffiltext{1}{Mizusawa VLBI Observatory, National Astronomical Observatory of Japan, \\
  2-12 Hoshigaoka-cho,  Mizusawa-ku, Oshu, Iwate 023-0861}
  \email{satoko.ss@nao.ac.jp}
 \altaffiltext{2}{Graduate school of Science and Engineering, Yamaguchi University, \\
  1677-1 Yoshida, Yamaguchi, Yamaguchi 753-8512}
\altaffiltext{3}{Shanghai Astronomical Observatory, Chinese Academy of Sciences, 
80 Nandan Road, Shanghai 200030, China}
\altaffiltext{4}{National Astronomical Observatory of Japan, 2-21-1 Osawa, Mitaka, Tokyo 181-8588}
\altaffiltext{5}{Department of Astronomy, 
Graduate University for Advanced Studies, 
2-21-1 Osawa, Mitaka, Tokyo 181-8588}

\maketitle

\begin{abstract}
We present the first internal motion measurement   
of the 6.7-GHz methanol maser within S269, 
a small H\emissiontype{II} region in the outer Galaxy,  
which was carried out in 2006 and 2011
using the Japanese VLBI Network (JVN).
Several maser groups and weak isolated spots were detected 
in an area spanning by $\sim$ 200 mas (1000 AU). 
Three remarkable maser groups are aligned 
at a position angle of $80^{\circ}$.
Two of three maser groups were also detected 
by a previous observation in 1998, 
which allowed us to study a long-term position variation of maser spots 
from 1998 to 2011. 
The angular separation between the two groups increased $\sim10$ mas, 
which corresponds to an expansion velocity of $\sim10$ km s$^{-1}$.
Some velocity gradient ($\sim10^{-2}$ km~s$^{-1}$ mas$^{-1}$) 
in the overall distribution was found. 
The internal motion between the maser groups support the hypothesis 
that the methanol masers in S269 could trace a bipolar outflow. 
\end{abstract}

\section{Introduction}

The $5_{1} \to 6_{0}  A^{+}$ methanol maser transition 
near 6.7 GHz discovered by Menten (1991) 
is known to be closely associated with high-mass star-forming regions
(e.g. \cite{caswell95}; \cite{walsh97}; \cite{minier03}; \cite{xu08}). 
They are thought to be a probe of just before and/or after the onset of   
the ultra compact (UC) H\emissiontype{II} region (e.g., \cite{walsh98}).  

Various morphologies of the 6.7 GHz methanol masers have been 
obtained with past interferometric and VLBI observations.  
The linear distribution and velocity gradient of the masers 
have often been explained as an edge-on rotation disk 
(e.g. \cite{norris98}).
Such a rotation disk hypothesis is supported by 
detections of internal motions surrounding a radio continuum 
source (\cite{sanna10a}; \cite{sanna10b}; \cite{goddi11}; \cite{moscadelli11}). 
Moreover, several ring-like distributions of masers have been applied to a disk model  
with rotation plus expansion/infall motion 
(e.g. \cite{bart05}; \cite{bart09}; \cite{sugiyama08b}; \cite{tor11}). 
However, De Buizer \etal (2012) resolved the near- and mid-infrared emission 
at the locations of four methanol maser rings, and
did not find consistent morphology of IR emission based on the hypothesis
that the masers reside in the circumstellar disks.

\begin{table*}
 \begin{center}
 \caption{Observation parameters}\label{tab:param}
 \begin{tabular}{cclccc}
  \hline \hline
Epoch & Date & \multicolumn{1}{c}{Telescopes\footnotemark[$*$]} & Synthesized beam & On-source time & $I_{\rm rms}$\footnotemark[$\dagger$] \\
        & [yyyy/mm/dd] &  & [mas, mas, degree]  & [hr] &  [mJy beam$^{-1}$] \\
  \hline
       1 & 2006/09/10 & Y, U, M, I & 23$\times$3, $-34$ & 2.1 & 110 \\
       2 & 2011/10/22 & Y, U, H, M, R, O, I & 8$\times$3, $-46$ & 6.5 & 10 \\ 
   \hline
  \multicolumn{5}{@{}l@{}}{\hbox to 0pt{\parbox{180mm}
{\footnotesize
\par\noindent
\footnotemark[$*$] Telescope code --- Y: Yamaguchi, U: Usuda, H: Hitachi, M: VERA-Mizusawa, R: VERA-Iriki, \\
O: VERA-Ogasawara, I: VERA-Ishigaki.
\par\noindent
\footnotemark[$\dagger$] Image rms noise level for line-free channel map.
}
 \hss}}
\end{tabular}
\end{center}
\end{table*}

On the other hand, a shock-wave model has been presented to
explain the maser location in several sources
(\cite{phillips98}; \cite{walsh98}; \cite{dodson04}). 
De Buizer (2003) searched for H$_2$ outflow signatures 
in massive young stellar objects with the linear distribution of 
methanol masers to test whether the outflows are
perpendicular to the linear distributions.  
Their search revealed that 
H$_2$ emission is distributed almost parallel to the distribution
of methanol masers in their sample sources. 
In addition, multi-epoch VLBI observations have shown the internal 
motions of methanol masers in massive star-forming region ON1,  
which suggested that the masers 
trace the expansion of the UC H\emissiontype{II} region 
or a bipolar outflow 
(\cite{rygl10}; \cite{sugiyama11}). 
The 6.7 GHz methanol maser emission has been detected 
on size scale of 1000 AU
from young stellar objects, 
and can be a powerful tool to investigate the environment close to the 
forming high-mass protostar. 
However, 
internal motions of the 6.7 GHz methanol masers
have been reported in limited number of sources so far  
(e.g. G16.59-0.05 for \cite{sanna10a}; G23.01-0.41 for \cite{sanna10b};
ON1 for \cite{rygl10}; \cite{sugiyama11}; 
IRAS 20126+4104 for \cite{moscadelli11}; 
W3(OH) for \cite{matsumoto11};
AFGL 5142 for \cite{goddi11}).

S269 (G196.45-1.68) 
is a small H\emissiontype{II} region spanning about 
2$^{\prime}$ 
(\cite{heydari82}), which is located in the outer Galaxy.
S269 harbors two bright near-infrared (IR) sources separated 
by $\sim30^{\prime\prime}$, 
IRS 1 and IRS 2 (\cite{wbn74}).  
Later, IRS 2 was resolved into double sources IRS 2e and IRS 2w 
with a separation of 4$^{\prime\prime}$ (\cite{eiroa94}; \cite{eiroa95}). 
Recent near-IR images imply that 
several H$_2$ knots are distributed across IRS2, 
which suggest two bipolar outflows,
powered by sources in IRS 2 (\cite{jiang03}).  
In S269, several signposts of star-forming activities such as 
OH and H$_2$O masers (e.g. \cite{www74}; \cite{genzel77}), 
high-velocity CO wing emission (\cite{wouterloot89}; \cite{yang02}) 
and a Herbig-Haro object (\cite{eiroa94})
have been detected.

The 6.7-GHz methanol maser emission has been also found  
in S269 (\cite{menten91}; \cite{szymczak00}).
The monitoring observations of 6.7-GHz methanol maser emission 
from 1999 to 2003 revealed that 
the three main methanol maser features have a sinusoidal time variation 
with a period of 668 days (\cite{goedhart04}). 
From past VLBI observation of 6.7-GHz methanol masers in S269 
using the European VLBI Network (EVN), 
Minier \etal (2000, hereafter M00) detected two maser groups  
named A and B; 
there is a single maser spot at velocity of 14.70 km~s$^{-1}$ in group A,
and eight maser spots at a velocity range of 15.04--15.43 km~s$^{-1}$ 
distributed in a linear structure of $\sim$15 mas 
at southeast--northwest direction with a clear velocity gradient in group B. 
The VLBI image by M00 showed that two groups were  
separated by $\sim55$ mas in 1998 November, 
and suggested that the methanol masers are probably associated 
with IRS 2. 
Recent astrometric observation of water maser in S269 
with VERA also revealed that 
the absolute position of the water maser feature agrees well 
with the position of IRS 2w (\cite{honma07}).

In this paper, we present 6.7-GHz methanol maser images of S269, 
eight and thirteen years after the past VLBI observation by M00, 
and report the first measurements of the internal motion of 
the S269 methanol maser. 
We adopt $D= 5.28$ kpc to S269 (\cite{honma07}), and 
hence 1 mas corresponds to 5.27 AU.

\section{Observations and Data Reduction}

VLBI observations of 6.7-GHz methanol masers in S269 were 
carried out at two epochs, on 2006 September 10 using four telescopes 
of the Japanese VLBI Network (JVN), and
on 2011 October 22 using seven telescopes (table~\ref{tab:param}).
At Epoch 1, left-circular polarization was received at Yamaguchi and Usuda, 
while linear polarization was received at VERA Mizusawa 
and VERA Ishigaki. 
At Epoch 2, left-circular polarization was received at all telescopes. 
The data were recorded with the VSOP terminal system 
(\cite{kawaguchi94}). 
The recorded data were correlated with the Mitaka FX correlator 
(\cite{shibata98}). 

Data reduction including calibration, data flagging, fringe fitting 
and imaging utilized using the NRAO AIPS package. 
For visibility-amplitude calibration, we applied the ``template method'' 
(e.g. \cite{diamond95}) in order to correct any short term gain fluctuations 
and pointing errors with a time resolution of 30 seconds. 

We corrected the visibility amplitude decrement due to 
the correlation between different (circular- or linear-) polarization 
data at Epoch 1 
under assumption of no significant polarization of the source,
using the method of Sugiyama \etal (2008a).  
A visibility amplitude in correlation between 
circular- and linear- polarization data reduces by $1/\sqrt{2}$. 
For the case of visibility amplitude in correlation 
between  linear- and linear- polarization data, 
the amplitude varies with time of observation, or parallactic 
angle of observing source at the time. 
We applied the amplitude correction factor for each baseline visibility. 

The residual delays and rates were estimated from 
observations of the continuum source 0611+131. 
The bandpass responses were calibrated with observations of 4C39.25.
Fringe fitting and self calibration were done for the brightest methanol maser emission in S269.
Channel maps were made every 0.178 km~s$^{-1}$ with 
uniform weighting. 
The positions of the detected maser spots were derived by 
fitting an elliptical Gaussian brightness distribution to the maps
using the AIPS task JMFIT. 
The observation parameters for the JVN observations are summarized 
in table~\ref{tab:param}.

We have also carried out the single-dish observations toward 
the 6.7-GHz methanol maser emission of S269 with the Yamaguchi 32-m telescope 
almost simultaneously with each JVN observation, 
for an absolute flux calibration to the VLBI data. 
The observations were performed for four days from 2006 September 4,
and one day on 2011 October 22 just after the JVN observations toward 
S269.

The dual circular polarizations were recorded simultaneously, 
and combined after being transformed into the spectra. 
The velocity resolution was 0.044 km~s$^{-1}$. 
The integration time was 14 minutes for each day, 
and the rms noise level was typically 1.0 Jy for one spectral channel.  
Spectral profile in September 2006 was obtained by 
averaging the four-day data, 
resulting the rms noise level of 0.5 Jy. 
Amplitude and gain calibrations were performed 
with the use of noise-sources having known noise temperatures.

\section{Results}

\begin{table}
\caption{Flux densities of the 6.7-GHz methanol maser peaks}\label{tab:peak}
  \begin{center}
    \begin{tabular}{ccc}
      \hline \hline
      V$_{\rm LSR}$ &   \multicolumn{2}{c}{Flux density} \\
      \cline{2-3}   
      & 2006 September & 2011 October \\ 
            \cline{2-3}
       [km s$^{-1}$] & \multicolumn{2}{c}{[Jy]}  \\
      \hline
       14.7  &  9.8 & 21.1 \\
       15.2  & 19.0 & 22.8 \\
       15.9  & 6.1 & 14.1 \\
   \hline
\end{tabular}
\end{center}
\end{table}

\subsection{Spectra}
\begin{figure}
  \begin{center}
     \FigureFile(80mm,140mm){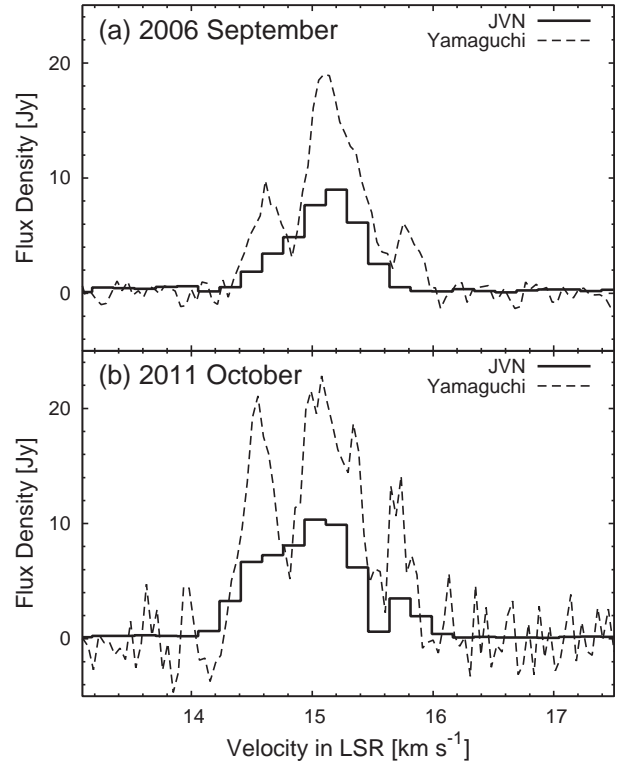}
  \end{center}
  \caption{
  Comparison of the cross-power spectrum (vector average) 
  obtained from 
  our JVN observations (Solid line) and the spectral profile measured 
  with the Yamaguchi 32-m telescope (Dashed line). 
Velocity resolution is 0.178 km~s$^{-1}$ and 0.044 km~s$^{-1}$,
respectively. 
 (a) Spectra in 2006 September.  
  The spectral profile measured with the Yamaguchi 32-m telescope  
  is obtained by averaging spectra observed from 2006 September 4 to 7. 
 (b) Spectra in 2011 October 22.
}
  \label{fig:spctr}
\end{figure}

Spectral profiles of the 6.7-GHz methanol maser emission in S269 
are shown in figure \ref{fig:spctr}. 
The maser emission was detected at velocity range of 14.0--16.5 km~s$^{-1}$. 
The spectrum detected with the Yamaguchi 32-m telescope 
revealed the brightest peak at velocity of 15.2 km~s$^{-1}$, 
and blue-shifted and red-shifted spectral components 
at peak velocities of 14.7 and 15.9 km~s$^{-1}$, 
which is consistent with the past single-dish observations 
(\cite{szymczak00}; \cite{goedhart04}). 
The flux densities at the peaks of 14.7, 15.2 and 15.9 km~s$^{-1}$ 
are summarized in table~\ref{tab:peak}. 
The values of the peak flux densities in 2006 lie 
within the variation range 
(20--50 Jy at 15.2 km~s$^{-1}$, 0--10 Jy at 14.7 and 15.9 km~s$^{-1}$)
observed during the period from 1999 January to 2003 March 
by single-dish monitoring (\cite{goedhart04}). 
In 2011, the peak flux densities at 14.7 and 15.9 km~s$^{-1}$ rise up 
over the variation range, 
while the peak value at 15.2 km~s$^{-1}$ is 
close to the value at its minimum intensity.  
This variation is different from the behavior 
during the period of the past monitoring observations.

\subsection{Spatial distributions}

\begin{figure*}
  \begin{center}
   \FigureFile(160mm,110mm){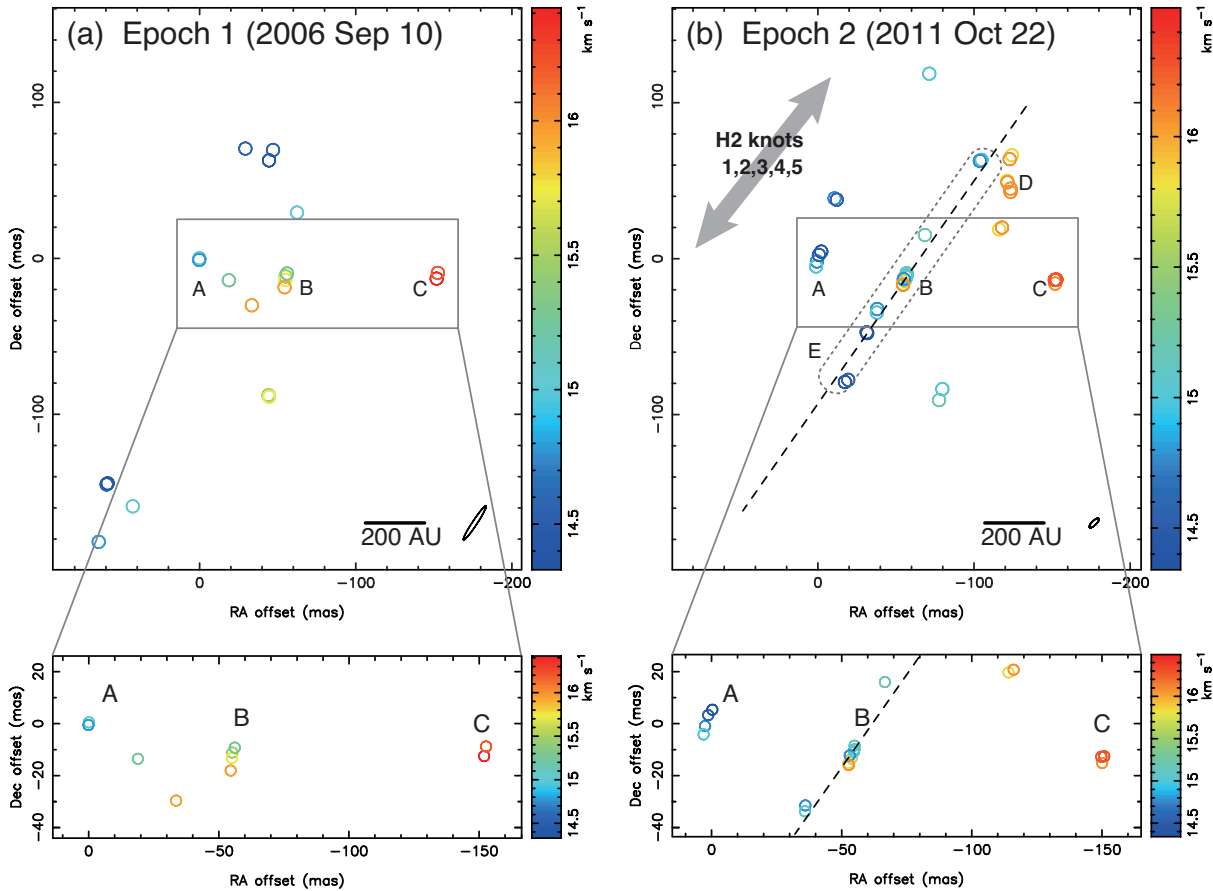}
  \end{center}
  \caption{
  The 6.7-GHz methanol maser distribution, observed 
  with the JVN on (a) 2006 September 10 and (b) 2011 October 22.  
  Color circles mark maser spots detected at the $>10 \sigma$ level
  in a single channel,
  or detected at the $>7 \sigma$ level in two adjacent channels 
  with the position overlapping within the beam size. 
  Color indicates the LSR doppler velocity of the spectral channel.
  The synthesized beam size is shown at the lower right in wide-field view. 
  In (b), 
  the dashed lines indicate the direction at position angle of 125$^{\circ}$, 
  and a grey arrow represents 
  the direction of alignment for the H$_2$ knots 
  1, 2, 3, 4 and 5 (\cite{jiang03}).  
    }
  \label{fig:image}
\end{figure*}

Figure \ref{fig:image} shows the 6.7-GHz methanol maser distribution 
in S269.  
At both of Epoch 1 and 2, 
all maser spots distribute over a range of $\sim200$ mas (1000 AU).
The maser distribution is organized by several maser groups 
and some weak isolated spots. 
Here, we give labels to remarkable groups (A, B, C, D and E). 
Groups A, B and C are seen at both Epoch 1 and 2, and 
aligned along the direction at a position angle of $\sim80^{\circ}$. 

The most luminous maser spot at 15.2 km~s$^{-1}$ belongs to group B. 
The maser spots in group B distribute 
in a linear structure of $\sim10$ mas 
in the southeast--northwest direction with a velocity gradient. 
The maser distribution of group B in 1998 by M00 show 
a similar linear structure of $\sim14$ mas 
in the southeast--northwest direction, 
and the size became smaller. 
The velocity ranges of detected maser spots in group B were similar
in 1998 and  2006 (Epoch 1);  
15.04 to 15.43 km~s$^{-1}$ in 1998 and 15.02 to 15.55 km~s$^{-1}$ in 2006. 
At Epoch 2, 
the velocity range of group B was broadened to 15.00--16.16 km~s$^{-1}$
and 
several spots at 15.67--16.16 km~s$^{-1}$ are newly detected. 
In 2011, the structure of group B is still elongated, 
but its velocity gradient became less simple.

Group A is the second brightest, which locates at 
$\sim55$ mas east from group B. 
At Epoch 1, it consists of two maser spots at velocities of 14.67 and 14.85 km~s$^{-1}$ 
with an angular separation of $\sim1$ mas. 
Its compact structure and velocity range agree well with the 
past VLBI observations in 1998 by M00.  
At Epoch 2, however, 
four spots with a velocity range of 14.84 to 15.33 km~s$^{-1}$ 
are linearly distributed over $\sim10$ mas with a clear velocity gradient.  

Group C is located 95 mas west from group B.   
It is divided into two spectral channels 
at velocities of 15.73 and 15.91 km~s$^{-1}$ at Epoch 1, 
while it consists of three maser spots 
at velocities of 16.33 and 16.49 km~s$^{-1}$ at Epoch 2. 
We cannot identify the detected maser spots within group C 
in the two epochs as being the same, 
because there is a difference of $\sim$0.5 km~s$^{-1}$ in velocity,  
larger than the velocity resolution of 0.178 km~s$^{-1}$, 
between two epochs in their LSR velocities. 

At Epoch 2, 
two prominent maser groups (labeled D and E in figure~\ref{fig:image}b
and table~\ref{tab:de}) are visible.  
Group D is seen at 70 mas west and 60 mas north 
from group B. 
Group D consists of several maser spots  
with velocities of 16.00 and 16.17 km~s$^{-1}$.  
The appearance of group D is consistent with brightening of the maser 
peak at 15.9 km~s$^{-1}$ in the spectral profile (figure~\ref{fig:spctr}b).
Group E consists of several blueshifted maser spots at velocities of 14.8--15.3 km~s$^{-1}$, 
and its detection is in accord with 
the brightening of the maser peak at 14.7 km~s$^{-1}$. 
They are framed in by a dotted rounded box in figure~\ref{fig:image}b, 
and distributed over $\sim170$ mas 
on the dashed line at a position angle of 125$^{\circ}$ in figure \ref{fig:image}b, 
almost parallel 
to the alignment direction of the H$_2$ knots 1,2,3,4 and 5. 
The maser spots in group B and E
seem to be aligned   
on the same line in the direction at position angle of 125$^{\circ}$. 
The two maser groups D and E are brighter than group C, but not 
as bright as group A. 

\begin{table*}
\caption{Parameters of maser spots in group D and E
at Epoch 2
}
\label{tab:de}
\begin{center}
   \begin{tabular}{cccc}
      \hline \hline
    V$_{\rm LSR}$ & Relative RA\footnotemark[$*$] & Relative Dec\footnotemark[$*$] & Peak Intensities \\
   \cline{2-3} 
  [km~s$^{-1}$] & \multicolumn{2}{c}{[mas]}  & [Jy beam$^{-1}$] \\
      \hline
      \multicolumn{4}{c}{Group D} \\
      \hline 
      16.00 & $-115.66$ & 18.78 & 0.50  \\
      16.00 & $-120.91$ & 49.77 & 0.38  \\
      16.00 & $-123.90$ & 66.23 & 0.32  \\
      16.16 & $-117.62$ & 19.86 & 0.32  \\
      16.16 & $-121.06$ & 48.82 & 0.29  \\
      16.16 & $-122.47$ & 63.70 & 0.26  \\
      16.16 & $-123.01$ & 42.72 & 0.27 \\
      16.16 & $-123.05$ & 44.84 & 0.26 \\
      \hline
      \multicolumn{4}{c}{Group E} \\
      \hline 
      14.84 &  $-17.03$ & $-79.03$ & 0.34 \\
      14.84 &  $-31.28$ & $-47.72$ & 0.33 \\
      15.00 &  $-19.16$ & $-77.65$ & 0.45 \\
      15.00 &  $-30.91$ & $-47.19$ & 0.46 \\
      15.17 &  $-37.65$ & $-32.30$ & 0.50 \\
      15.17 & $-120.91$ & $49.77$ & 0.38 \\
      15.33 &  $-37.48$ & $-34.57$ & 0.47 \\
      15.33 & $-104.48$ & $63.27$ & 0.41 \\
      15.50 &  $-68.20$ &  $15.10$ & 0.44 \\
      \hline 
\multicolumn{4}{@{}l@{}}{\hbox to 0pt{\parbox{85mm}{\footnotesize
  \footnotemark[$*$] Relative position with respect to the barycentric point of 
  group A.
  \par\noindent
     }
     \hss}}
    \end{tabular}
  \end{center}
 \end{table*}

The maser spots, except groups A and B were not visible   
in the EVN observations in 1998 November (M00). 
The maser peaks at 14.7 and 15.9 km~s$^{-1}$ could be 
at the minimum phase in intensity at that time, 
as Goedhart et al. (2004) has shown.

\subsection{Internal motion}

\begin{figure*}
  \begin{center}
    \FigureFile(160mm,80mm){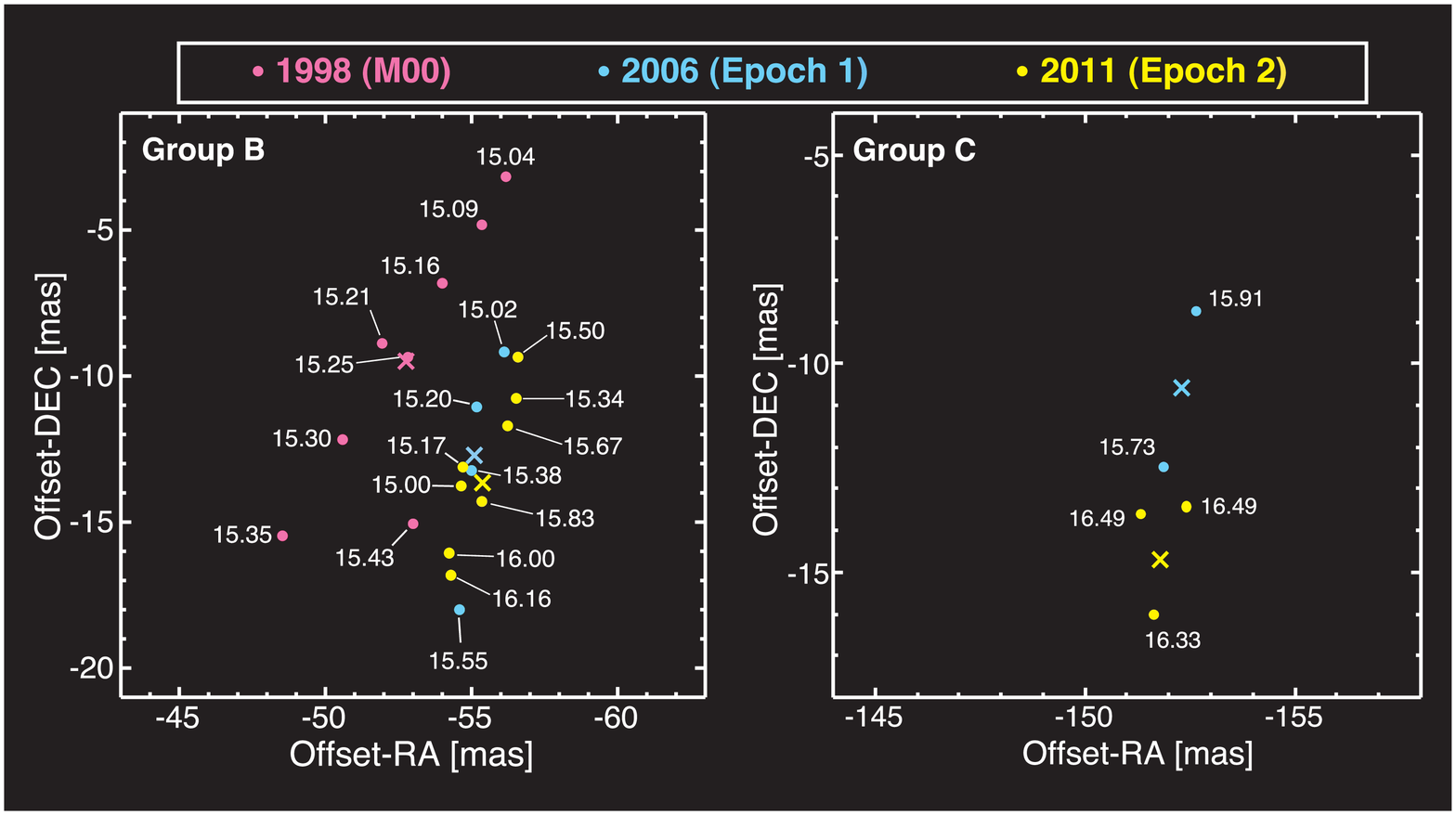}
  \end{center}
  \caption{
 Superposed maps of the 6.7-GHz methanol maser distribution 
 in groups B and C with respect to group A 
 in 1998 (M00), 2006 (Epoch 1) and 2011 (Epoch 2).
 Color circles mark the maser spots, and a  
 cross represents the barycentric point for each group and epoch. 
 The coordinate origin (0,0) is set to the barycentric point among maser spots 
 within group A for each epoch. 
For 1998, the origin (0,0) is the position of the single maser spot 
at a velocity of 14.70 km~s$^{-1}$ in group A. 
  }
  \label{fig:evnjvn}
\end{figure*}

Since we did not use the technique of phase referencing 
for our VLBI observations, 
the absolute information of position is unknown. 
Furthermore, there is uncertainty about the identification of each maser spot,  
because (i) the different velocity resolutions between EVN and JVN; 
and (ii) time variation of velocity ranges of groups A, B and C at Epoch 2.   
Therefore, we here describe the relative position of the maser spots
with respect to  
the barycentric point among maser spots within group A for each epoch. 
For 1998, the reference point is the position of the single maser spot 
at a velocity of 14.70 km~s$^{-1}$ in group A.

The relative positions of the maser spots in groups A, B and C 
are listed in table~\ref{tab:maserspots}.
The superposition of the maser distributions in 1998 (M00), 
2006 (Epoch 1) and 2011 (Epoch 2) is shown in figure~\ref{fig:evnjvn}.
The spatial and velocity structure of groups A, B are similar 
during the period from 1998 to 2011.
Group C appears in 2006 (Epoch 1) and 2011 (Epoch 2), 
and there are similarities concerning the position and the compactness of its structure
between the two epochs.
The angular separation between the barycentric points of groups A and B 
increases 5 mas 
for thirteen yeas from 1998 to 2011.
If we assume that it increases at a constant rate, 
the velocity of the internal motion is estimated to be 10 km~s$^{-1}$. 
We note that the maser distribution and the velocity range of group A 
at Epoch 2
are changed from two previous VLBI observations 
in 1998 and 2006 (Epoch 1),  
and there could be uncertainty about the reference point at Epoch 2. 
If we exclude the results at Epoch 2 for estimation of the internal motion 
between group A and B, 
the velocity of the motion would be calculated to be 13 km~s$^{-1}$, 
since the angular separation increases by 4 mas from 1998 to 2006.

\begin{table}
\caption{Positions of the maser spots in group A, B and C}\label{tab:maserspots}
\begin{center}
   \begin{tabular}{cccc}
      \hline \hline
      Group & V$_{\rm LSR}$ & Relative RA\footnotemark[$*$] & Relative Dec\footnotemark[$*$]  \\
   \cline{3-4} 
  & [km~s$^{-1}$] & \multicolumn{2}{c}{[mas]}  \\
      \hline
      \multicolumn{4}{c}{1998 (M00)} \\
      \hline 
      A & 14.70 & 0.00 & 0.00  \\
      B & 15.04 & $-56.15$ & $-3.17$  \\
      B & 15.09 & $-55.35$ & $-4.89$  \\
      B & 15.16 & $-54.01$ & $-6.81$  \\
      B & 15.21 & $-51.93$ & $-8.89$  \\
      B & 15.25 & $-52.80$ & $-9.35$  \\
      B & 15.30 & $-50.59$ & $-12.16$ \\
      B & 15.35 & $-48.54$ & $-15.47$ \\
      B & 15.43 & $-53.00$ & $-15.03$ \\
      \hline
      \multicolumn{4}{c}{2006 (Epoch 1)} \\
      \hline 
      A & 14.67 &    0.08    & $-0.51$ \\
      A & 14.85 & $-0.08$ & 0.51 \\
      B & 15.02 & $-56.13$ & $-9.18$ \\
      B & 15.20 & $-55.19$ & $-11.06$ \\
      B & 15.38 & $-54.99$ & $-13.20$ \\
      B & 15.55 & $-54.57$ & $-18.00$ \\
      C & 15.73 & $-151.88$ & $-12.46$ \\
      C & 15.91 & $-152.64$ & $-8.75$ \\
      \hline
      \multicolumn{4}{c}{2011 (Epoch 2)} \\
      \hline 
      A & 14.84 & $-1.93$ & 4.47 \\
      A & 15.00 & $-0.35$ & 2.38 \\
      A & 15.17 & 0.86 & $-1.81$ \\
      A & 15.34 & 1.43 & $-5.04$ \\
      B & 15.00 & $-54.65$ & $-13.74$ \\
      B & 15.17 & $-54.68$ & $-13.10$ \\
      B & 15.34 & $-56.29$ & $-11.65$ \\
      B & 15.50 & $-56.56$ & $-9.34$  \\
      B & 15.67 & $-56.24$ & $-11.68$ \\
      B & 15.83 & $-55.36$ & $-14.27$ \\
      B & 16.00 & $-54.22$ & $-16.08$ \\
      B & 16.16 & $-54.29$ & $-16.78$ \\
      C & 16.33 & $-152.42$ & $-13.42$ \\	
      C & 16.49 & $-151.31$ & $-13.60$ \\	
      C & 16.49 & $-151.62$ & $-16.01$ \\	
\hline 
\multicolumn{4}{@{}l@{}}{\hbox to 0pt{\parbox{85mm}{\footnotesize
  \footnotemark[$*$] {\scriptsize Relative position with respect to the barycentric point of 
  group A.}
  \par\noindent
     }\hss}}
    \end{tabular}
  \end{center}
 \end{table}

It is difficult to discuss about the motion of group C,  
because the velocity range of group C at Epoch 1 and 2 is not same, and 
therefore the maser spots in group C in different epochs 
cannot be identified as being the same.

\subsection{Velocity gradient}

\begin{figure}
  \begin{center}
   \FigureFile(80mm,60mm){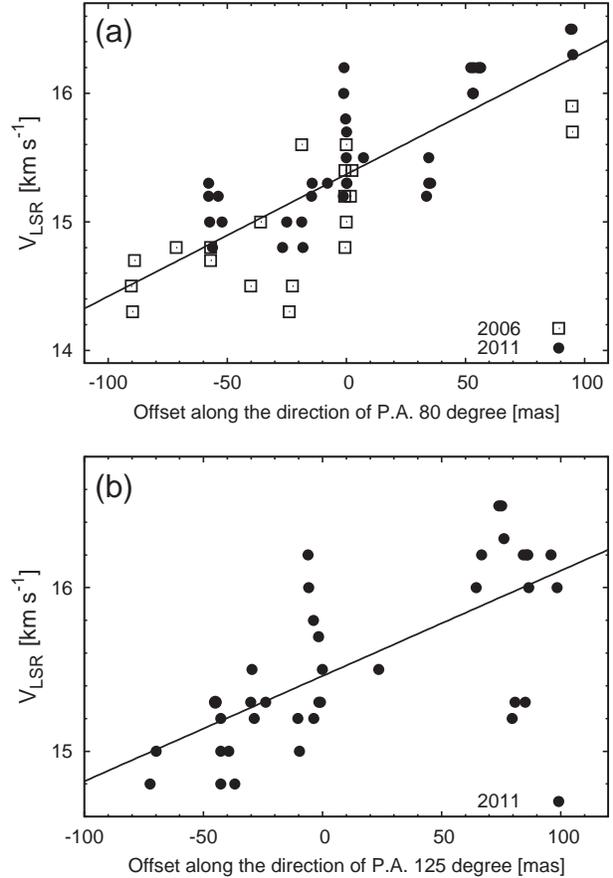}
  \end{center}
  \caption{
  (a) Position-Velocity diagram of all the maser spots detected in 2006 and 2011,
  along the direction at position angle of 80$^{\circ}$.
  The open square and 
  the filled circle indicates the maser spots obtained from 
  VLBI observations in 2006 and 2011, respectively. 
  The solid line represents a linear fit and indicates a velocity gradient of 
  $9.5$ m~s$^{-1}$~mas$^{-1}$.
  (b) Position-Velocity diagram of all the maser spots detected in 2011 
  along the direction at position angle of 125$^{\circ}$.
  The solid line represents a linear fit and indicates a velocity gradient of 
  $6.4$ m~s$^{-1}$~mas$^{-1}$.
    }
  \label{fig:velgra}
\end{figure}

The velocity gradient in the overall distribution could roughly be seen 
along the alignment of groups A, B and C  
at a position angle of $\sim80^{\circ}$.  
At Epoch 2, another obvious linear structure, which consists of 
groups B and E, is seen 
along the southeast--northwest direction (position angle of 125$^{\circ}$) 
in figure~\ref{fig:image}. 
However, a more rigorous inspection for the individual maser groups
indicates that the velocity gradient is not simple. 
A position-velocity diagram  
along the direction at position angles of $80^{\circ}$ and 125$^{\circ}$ 
indicates a linear fitting of velocity gradient of 
9.5 m~s$^{-1}$~mas$^{-1}$ and 
6.4 m~s$^{-1}$~mas$^{-1}$, 
respectively
(figure~\ref{fig:velgra}).

\section{Discussions}

As described in introduction, 
the 6.7-GHz methanol maser emission has been considered to trace 
various structures in high-mass star-forming region. 
Here, we discuss possible scenarios 
where 6.7-GHz methanol maser emission is 
associated with a bipolar outflow or a disk. 

\subsection{Outflow scenario}

\begin{figure}
  \begin{center}
      \FigureFile(80mm,60mm){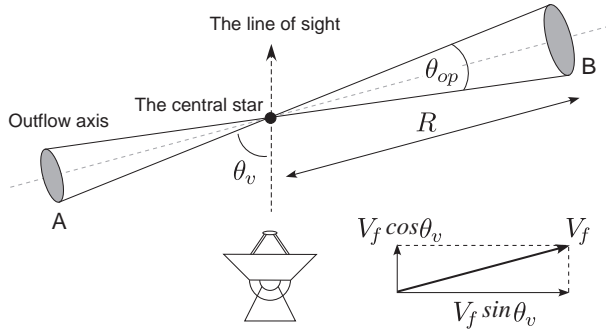}
  \end{center}
  \caption{
 Possible schematic diagram of outflow model 
 to explain the increase of the angular separation between two maser 
 groups,  A and B. 
Groups A and B travel away 
 at the outflow velocity of $V_f$ from the central star. 
  }
  \label{fig:outflow}
\end{figure}

The bipolar outflow scenario is the simplest explanation for the increase  
of the angular separation between two maser groups A and B. 
The axis of outflow would be a line joining the two maser groups A and B,
in the direction at position angle of $\sim80^{\circ}$. 
A possible schematic diagram of the outflow model is shown 
in figure~\ref{fig:outflow}. 
The viewing angle of the outflow ($\theta_{v}$) is estimated to be 87$^{\circ}$, 
from the projected relative velocity of 13 km~s$^{-1}$ 
($V_f \sin{\theta_v}$)
and the velocity difference between groups A and B 
of 0.6 km~s$^{-1}$ ($V_f \cos{\theta_v}$). 
Therefore, the outflow axis is almost parallel to the sky plane, 
and the absolute velocity of the expansion between groups A and B 
is estimated to be nearly 13 km~s$^{-1}$. 
We derived the momentum rate of the outflow $\dot{P}_f$, defined as 
\begin{equation}
\dot{P}_f= 2 \Omega_f R^2 n_{{\rm H}_2} m_{{\rm H}_2} V_f^2, 
\end{equation}
where $\Omega_f$ is the solid angle of the outflow,  
$R$ is the distance from the central star, 
$n_{{\rm H}_2}$ and $m_{{\rm H}_2}$ are the number density,  
and the weight of molecular hydrogen, and 
$V_f$ is the outflow velocity (e.g. \cite{nagayama08}).
The location of the central star would be somewhere on the line segment 
between groups A and B. 
The solid angle of the outflow, $\Omega_f$ is 
$2\pi\{1-\cos(\theta_{op}/2)\}$, 
and 
the opening angle of the outflow, $\theta_{op}$,  is an 
appearance angle of group A or B from the central star.
Here, we assume that the central star is located at the midpoint between the 
groups A and B.  
We obtain $\dot{P}_f = 2.1 \times 10^{-5} \MO$ km s$^{-1}$ yr$^{-1}$, 
adopting $\theta_{op}=10^{\circ}$ as the appearance angle of 
group A or B from the midpoint that 
$V_f=$ 6.5 km~s$^{-1}$ as half of the absolute velocity of 
the expansion between groups A and B, 
$R= 150$ AU as half of the distance between groups A and B, 
and a typical gas density for methanol maser environment 
$n_{{\rm H}_2}=10^8$ cm$^{-3}$ (\cite{cragg05}).

Group C is also located on the extension of the line, 
and groups A, B and C could be associated with the same outflow.  
However, the direction at a position angle of $\sim80^{\circ}$ 
is not consistent with a large-scale bipolar outflow 
in the southeast--northwest direction traced by several H$_2$ knots 
(Knots 1,2,3,4,5; \cite{jiang03}). 
Therefore, the internal motion of groups A and B could be driven 
by another outflow. 
The near-IR images have suggested the existence of a second outflow 
with a different axis 
traced by H$_2$ knot 6 as well (\cite{jiang03}). 
If the second outflow is powered by sources in IRS 2, 
the outflow axis would be a line to connect IRS 2 and H$_2$ knot 6.
We note that the direction of the outflow axis is nearly parallel 
to the position angle of $80^{\circ}$, 
which is the alignment of maser groups A, B and C. 
Currently, there is no strong observational support to link 
the outflow in southeast--northwest direction traced by several H$_2$ knots 
with the maser spots in group B and E aligned 
along the direction at position angle of $125^{\circ}$ seen at Epoch 2. 
Measurements of the absolute positions and proper motions of 
the methanol masers are necessary 
in order to confirm the relation 
between the maser spots and the H$_2$ knots 
in southeast--northwest direction.

\subsection{Disk scenario}

\begin{figure}
  \begin{center}
    \FigureFile(80mm,60mm){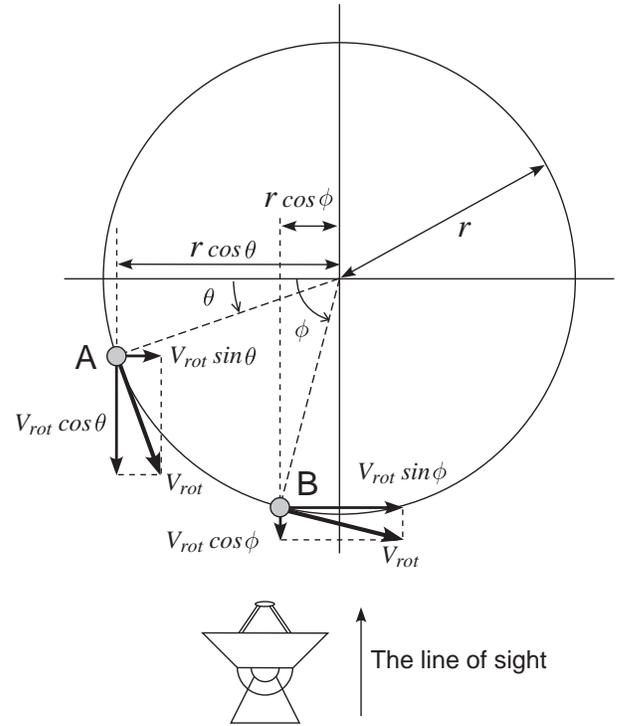}
  \end{center}
  \caption{
 Diagram of a rotation disk from the face-on view. 
 Here, we assume that the maser groups A and B rotate in an orbit with 
 the radius of $r$ at the rotation velocity of $V_{rot}$.
  }
  \label{fig:disk}
\end{figure}

The disk scenario has often been proposed 
to interpret the maser distribution 
in a linear distribution with a velocity gradient, 
and could be another candidate to explain the observational results 
of the methanol masers.
If the velocity gradient of the methanol maser emission is 
due to a rotation disk, 
the enclosed mass in the Keplerian law is estimated to be $\sim1 \MO$, 
assuming an edge-on view of the disk with a disk radius of 1000 AU,
the total extent of the methanol masers. 
The estimated value is too small as a high-mass star forming region, 
in which 6.7-GHz methanol maser emission is exhibited. 
This fact has already been pointed out by M00, 
which suggested the possibility that methanol masers are associated with  
a fraction of the rotation disk, 
resulting is an underestimate of the enclosed mass. 

Here, we consider the hypothesis that 
the methanol masers could trace the partial disk, 
assuming that groups A and B rotate around a protostar 
in an orbit 
with a radius of $r$ at a rotation velocity of $V_{rot}$, 
as shown in figure~\ref{fig:disk}. 
The projected distance between groups A and B 
($D_{AB}=r \cos{\theta} - r \cos{\phi}$), 
the velocity of the relative motion between groups A and B 
($V_{rel}=V_{rot} \sin{\phi} - V_{rot} \sin{\theta}$), 
and 
the difference of the radial velocity between groups A and B 
($\Delta V_{rad}=V_{rot} \cos{\theta} - V_{rot} \cos{\phi}$)
are provided from our VLBI observations.
If we assume that groups A and B are located in the near side 
of the disk  
($0^{\circ} \leq \theta \leq 180^{\circ}$ 
and $0^{\circ} \leq \phi \leq 180^{\circ}$), 
the orbit radius and the rotation velocity are estimated to be 
$r \sim 75000$  AU and $V_{rot} \sim 150$ km~s$^{-1}$, 
adopting the observational results: $D_{AB}$ of 300 AU, 
$V_{rel}$ of 13 km~s$^{-1}$ and $\Delta V_{rad}$ of 0.6 km~s$^{-1}$. 
Using the estimated $r$ and $V_{rot}$, 
the resulting enclosed mass is $1.9 \times 10^6 \MO$, 
which makes this scenario unlikely.
If the location of group A is allowed to be at the far side of the disk
($-180^{\circ} \leq \theta \leq 0^{\circ}$), 
much smaller values of the enclosed mass can be derived. 
When $(\theta, \phi)$ is $(-88^{\circ}, 92^{\circ})$, 
the minimum value of the enclosed mass is calculated to be $180 \MO$, 
in a disk property with $V_{rot}=6.5$ km~s$^{-1}$ and $r = 3300$ AU. 
The enclosed mass of $\geq180 \MO$ is still somewhat large 
as a single high-mass protostar. 
Therefore, we conclude that the scenario that 
methanol masers trace a disk around a protostar is less likely 
to account for the observed characteristics of the methanol maser emission.

\vspace{5mm}
We acknowledge the referee, Anna Bartkiewicz, for careful reading and 
her constructive suggestions. 
We also express our appreciation to Vincent Minier for informing us of his results. 
The JVN project is led by the National Astronomical Observatory of Japan (NAOJ), 
which is a branch of the National Institutes of Natural Sciences (NINS), 
Hokkaido University, Ibaraki University, University of Tsukuba, 
Gifu University, Osaka Prefecture University, Yamaguchi University, 
and Kagoshima University, 
in cooperation with the Geographical Survey Institute (GSI), 
the Japan Aerospace Exploration Agency (JAXA), 
and the National Institute of Information and Communications Technology (NICT).
We are grateful to all staff members of JVN for supporting our observations.


\begin{thebibliography}{}
\bibitem[Bartkiewicz \etal (2005)]{bart05}
  Bartkiewicz, A., Szymczak, M.,  van Langevelde, H. J. \ 2005, \aap,
442, L61
\bibitem[Bartkiewicz \etal (2009)]{bart09}
  Bartkiewicz, A., Szymczak, M.,  van Langevelde, H. J.,  
  Richards, A. M. S.; Pihlstr$\ddot{\rm o}$m, Y. M.
  \ 2009, \aap, 502, 155
\bibitem[Cragg \etal (2005)]{cragg05}
 Cragg, D. M., Sobolev, A. M., Godfrey, P. D. \ 2005, \mnras, 360, 533
\bibitem[Caswell \etal (1995)]{caswell95}
  Caswell, J. L., Vaile, R. A., Ellingsen, S. P., Whiteoak, J. B., Norris, R. P.
  \ 1995, \mnras, 272, 96
\bibitem[De Buizer (2003)]{debuizer03}
De Buizer, J. M. \ 2003, \mnras, 341, 277
\bibitem[De Buizer \etal (2012)]{debuizer12} De Buizer, J.~M., 
Bartkiewicz, A., \& Szymczak, M.\ 2012, \apj, 754, 149 
\bibitem[Diamond (1995)]{diamond95}
   Diamond, P.J., \ 1995, in ASP Conf. Ser. 82, Very Long Baseline Interferometry and the VLBA, ed. J.A. Zensus, P.J. Diamond, 
   $\&$ P.J. Napier (San Francisco: ASP), 227
\bibitem[Dodson \etal (2004)]{dodson04}
   Dodson, R., Ojha, R., Ellingsen, S. P. \ 2004, \mnras, 351, 779
\bibitem[Eiroa \etal (1994)]{eiroa94}
   Eiroa, C., Casali, M. M., Miranda, L. F., Ortiz, E. \ 1994, 
   \aap, 290, 599
\bibitem[Eiroa \& Casali (1995)]{eiroa95}
   Eiroa, C., Casali, M. M. \ 1995, \aap, 303, 87
\bibitem[Genzel \& Downes (1977)]{genzel77}
   Genzel, R., Downes, D., 1977, \aaps, 30, 145
\bibitem[Goddi \etal (2011)]{goddi11}
   Goddi, C., Moscadelli, L., Sanna, A., 2011, \aap, 535, L8 
\bibitem[Goedhart \etal (2004)]{goedhart04}
   Goedhart, S., Gaylard, M. J., van der Walt, D. J. \ 2004, 
   \mnras, 355, 553
\bibitem[Heydari-Malayeri \etal (1982)]{heydari82}
   Heydari-Malayeri, M., Testor, G., Baudry, A., Lafon, G., de La Noe, J., \ 1982,
   \aap, 113, 118
\bibitem[Honma \etal (2007)]{honma07}
   Honma, M. \etal \ 2007, \pasj, 59, 88
\bibitem[Jiang \etal (2003)]{jiang03}
   Jiang, Z. \etal  \ 2003,    \apj,  596, 1064
\bibitem[Kawaguchi \etal (1994)]{kawaguchi94}
   Kawaguchi, N., Kobayashi, H., Miyaji, T., Mikoshiba, H., Tojo, A., Yamamoto, Z., Hirosawa, H. 1994, in VLBI Technology Progress and Future Observational Possibilities, ed. T. Sasao, S. Manabe, O. Kameya, M. Inoue (Tokyo: Terra Scientific Publishing Company), 26
\bibitem[Matsumoto \etal (2011)]{matsumoto11}
  Matsumoto, N., Honma, M., Isono, Y., Ujihara, H., Kimura, K.,  
  Matsumoto, K., Sawada-Satoh, S., Doi, A., Fujisawa, K., Ueno, Y., 
  \pasj, 63, 1345
\bibitem[Menten (1991)]{menten91}
   Menten, K. M. \ 1991, \apj, 380, L75
\bibitem[Minier \etal (2000)]{minier00}
   Minier, V., Booth, R. S., Conway, J. E.  \ 2000, \aap, 362, 1093 (M00)
\bibitem[Minier \etal (2003)]{minier03}
   Minier, V., Ellingsen, S. P., Norris, R. P., Booth, R. S. \ 2003, 
   \aap, 403, 1095
\bibitem[Moscadelli \etal (2011)]{moscadelli11}
   Moscadelli, L.; Cesaroni, R.; Rioja, M. J.; Dodson, R.; Reid, M. J. \ 2011, 
   \aap, 526, A66
\bibitem[Nagayama \etal (2008)]{nagayama08}
  Nagayama, T. \etal \ 2008
\pasj, 60, 1069
\bibitem[Norris \etal (1998)]{norris98}
   Norris, R. P., Byleveld, S. E., Diamond, P. J., Ellingsen, S. P.,
   Ferris, R. H., Gough, R. G., Kesteven, M. J., McCulloch, P. M.,
   Phillips, C. J., Reynolds, J. E., Tzioumis, A. K., Takahashi, Y., 
   Troup, E. R., Wellington, K. J.
   \ 1998, \apj, 508, 275
\bibitem[Phillips \etal (1998)]{phillips98}
   Phillips, C. J., Norris, R. P., Ellingsen, S. P., McCulloch, P. M. 
   \ 1998, \mnras, 300, 1131
\bibitem[Rygl \etal (2010)]{rygl10}
  Rygl, K. L. J., Brunthaler, A., Reid, M. J., Menten, K. M., 
  van Langevelde, H. J., Xu, Y. \ 2010, \aap, 511, A2
\bibitem[Sanna \etal (2010a)]{sanna10a}
   Sanna, A., Moscadelli, L., Cesaroni, R., Tarchi, A., Furuya, R. S., 
   Goddi, C. \ 2010a, \aap, 517, A71
\bibitem[Sanna \etal (2010b)]{sanna10b}
   Sanna, A., Moscadelli, L., Cesaroni, R., Tarchi, A., Furuya, R. S., 
   Goddi, C. \ 2010b, \aap, 517, A78
\bibitem[Shibata \etal (1998)]{shibata98}
Shibata, K. M., Kameno, S., Inoue, M., Kobayashi, H.
\ 1998, in Radio Emission from Galactic and Extragalactic Compact Sources, 
ASP Conference Series, Volume 144, 
IAU Colloquium 164, eds. J.A. Zensus, G.B. Taylor, \& J.M. Wrobel p. 413.
\bibitem[Sugiyama \etal (2008a)]{sugiyama08a}
  Sugiyama, K., Fujisawa, K., Doi, A., Honma, M., Kobayashi, H., 
  Bushimata, T., Mochizuki, N., Murata, Y.  
  \ 2008, \pasj, 60, 23
\bibitem[Sugiyama \etal (2008b)]{sugiyama08b} 
   Sugiyama, K., Fujisawa, K., Doi, A., Honma, M., Isono, Y., Kobayashi, H., 
   Mochizuki, N.,  Murata, Y.  \ 2008, \pasj, 60, 1001
\bibitem[Sugiyama \etal (2011)]{sugiyama11} 
   Sugiyama, K., Fujisawa, K., Doi, A., Honma, M., Isono, Y., Kobayashi, H., Mochizuki, N., 
   Murata, Y., Sawada-Satoh, S., Wajima, K. 
   \ 2011, \pasj, 63, 53
\bibitem[Szymczak, Hrynek \& Kus (2000)]{szymczak00}
   Szymczak, M., Hrynek, G., Kus, A. J. \ 2000, 
   \aaps, 143, 269
\bibitem[Torstensson \etal (2011)]{tor11}
   Torstensson, K. J. E.,  van Langevelde, H. J.,  Vlemmings, W. H. T.,
   Bourke, S. \ 2011, \aap, 526, A38
\bibitem[Walsh \etal (1997)]{walsh97}
   Walsh, A. J., Hyland, A. R., Robinson, G.,  Burton, M. G. \ 1997, 
   \mnras, 291, 261
 \bibitem[Walsh \etal (1998)]{walsh98}
   Walsh, A. J., Burton, M. G., Hyland, A. R., Robinson, G. \ 1998,
   \mnras, 301, 640 
\bibitem[Wouterloot \& Brand (1989)]{wouterloot89}
  Wouterloot, J.G.A., Brand, J., 1989, \aaps, 80. 149
\bibitem[Wynn-Williams \etal (1974a)]{wbn74}
   Wynn-Williams, C. G., Becklin, E. E., Neugebauer, G. \ 1974a, 
   \apj, 187,  473
\bibitem[Wynn-Williams \etal (1974b)]{www74}
   Wynn-Williams, C. G., Werner, M. W., Wilson, W. J. \ 1974b, 
   \apj, 187,  41
\bibitem[Xu \etal (2008)]{xu08}
   Xu, Y., Li, J. J., Hachisuka, K., Pandian, J. D., Menten, K. M.,
    Henkel, C. \ 2008, \aap, 485, 729
\bibitem[Yang \etal (2002)]{yang02}
  Yang, J., Jiang, Z., Wang, M., Ju, B., Wang, H. \ 2002, \apjs, 141, 157
\end{thebibliography}
\end{document}